\documentclass[11pt,nofootinbib,floatfix]{revtex4}
\usepackage{mathrsfs}
\usepackage{graphicx}
\usepackage{amsmath}
\usepackage{amsfonts}
\usepackage{amssymb}
\usepackage{color}

\usepackage{epsfig}
\usepackage{CJK}
\usepackage{eepic}
\usepackage{bbm}
\usepackage{dcolumn}
\usepackage{bm}
\usepackage{ulem}
\usepackage{slashbox}
\usepackage{multirow}

\newcommand{\omits}[1]{}

\def\bc{\begin{center}}
\def\nno{\nonumber}
\def\ec{\end{center}}
\def\be{\begin{eqnarray}}
\def\ee{\end{eqnarray}}

\newcommand{\nc}{\newcommand}
\nc{\rnc}{\renewcommand} \nc{\ket}[1]{\left | \, #1 \right \rangle}
\nc{\bra}[1]{\left \langle #1 \, \right |}
\nc{\ua}{\uparrow} \nc{\da}{\downarrow}

\nc{\braket}[2]{\langle\, #1\,|\,#2\,\rangle}
\nc{\half}{\frac{1}{2}}

\nc{\prj}{\mathcal{P}} \nc{\hilb}{\mathcal{H}}
\nc{\pth}{\mathcal{C}} \nc{\inprod}[2]{\braket{#1}{#2}}
\nc{\upket}{\ket{\uparrow}} \nc{\downket}{\ket{\downarrow}}
\nc{\upbra}{\bra{\uparrow}} \nc{\downbra}{\bra{\downarrow}}

\begin{document}


\title{On the emergence of gravitational dynamics from tensor networks}

\author{Jia-Rui Sun$^{1}$} \email{sunjiarui@mail.sysu.edu.cn}
\author{Yuan Sun$^{1}$} \email{sunyuan6@mail.sysu.edu.cn}

\affiliation{${}^1$School of Physics and Astronomy, Sun Yat-Sen University, Guangzhou 510275, China}



\begin{abstract}
Tensor networks are powerful techniques that widely used in condensed matter physics. In this language, the wave function of a quantum manybody system is described by a network of tensors with specific entanglement structures. Recently, it is shown that tensor network can generate the anti-de Sitter (AdS) geometry by using the entanglement renormalization approach. However, whether the dynamical connections can be found between the tensor network and the gravity is an important unsolved problem. In this paper, we give a novel proposal to integrate ideas from tensor networks, entanglement entropy, canonical quantization of quantum gravity and the holographic principle and argue that the gravitational dynamics can be generated from a tensor network if the wave function of the latter satisfies the Wheeler-DeWitt equation.

\end{abstract}


\maketitle

\newpage

\section{Introduction}
In the study of quantum manybody physics, tensor network becomes a natural language in which the wave function of the system is described by a series of tensors which comprise into a network, each tensor can be viewed as a building block of the wave function and the connections between tensors are captured by quantum entanglement among the particles. Typically, the total Hilbert space of a quantum manybody system is too large to handle with due to the large number of particles and their microstates. Efficient ways to deal with the problem is to utilize the idea of real space renormalization group (RG) to make the number of coarse grained effective degrees of freedom (d.o.f.) reduce dramatically. The RG approach for tensor network is called the multi-scale entanglement renormalization, which was shown very powerful both for theoretical and numerical calculations~\cite{Vidal:2007hda}.

Recently, an interesting progress was that the entanglement renormalization of tensor networks can be viewed as a discrete version of the AdS/CFT correspondence. More specifically, a discrete time slice of AdS geometry can emerge from the coarse graining of some tensor network at the quantum critical point~\cite{Swingle:2009bg}. Soon after, different kinds of tensor networks have been investigated to generate the AdS geometry, with the attempt to construct the bulk spacetime (or gravity) by using the informations of the boundary quantum theory such as the correlation functions and entanglement entropy~\cite{Pastawski:2015qua,Hayden:2016cfa}.

Indeed, in the construction of bulk geometry, quantum entanglement or entanglement entropy was shown to play a vital important role. Previous evidences include the proposal of calculating entanglement entropy of the boundary CFT from minimal surface in bulk AdS~\cite{Ryu:2006bv}, the proposal that spacetime can emerge from the quantum entanglement of boundary CFT, in which disentangling CFTs in two boundary regions can make the bulk spacetime disconnected~\cite{VanRaamsdonk:2010pw}, the study of rebuilding bulk AdS geometry from the entanglement wedge of the boundary CFT~\cite{Dong:2016eik} and so on. Among these approaches, the essential point is to find out the underlying connections between the dynamics of the non-gravitational system (such as the CFT) and that of the spacetime geometry, i.e. the gravitational dynamics. Otherwise, the geometries emerged from the non-gravitational systems are only an analogy. Similar situations occurred in the study of analogue gravity~\cite{Barcelo:2005fc}, it was not until recently that the dynamical connections between acoustic black holes (one kind of the analogue gravity) and the real black holes have been revealed~\cite{Ge:2015uaa,Sun:2017eph} .

As for the tensor network approach of building spacetime geometry is concerned, the crucial question is whether the gravitational dynamics, namely, Einstein's equation can also be constructed (or generated) from the tensor network, and hence, from the non-gravitational quantum manybody systems. In this paper, we present a novel proposal to combine the key ideas from the tensor networks, entanglement entropy, canonical quantization of quantum gravity and holographic principle together and argue that Einstein's equation can be generated from the tensor network if the Schr\"{o}dinger equation which satisfied by the wave function of the tensor network can be rewritten as the Wheeler-DeWitt equation.

\section{Equivalence between the wave functions}
Considering a quantum manybody system (with $N$ particles) in $d$-dimensional flat spacetime, its ground state wave function $|\Psi\rangle$ can be
expressed as
\be\label{GSWF1}
|\Psi\rangle=\sum_{a_1\cdots a_N}T_{a_1\cdots a_N}|a_1\rangle\otimes\cdots\otimes|a_N\rangle,
\ee
where $|a_j\rangle$ is the basis of the $j$-th particle, and $T_{a_1\cdots a_N}$ can be viewed as the coefficients of a $N$-rank tensor. In the tensor network representation, the tensor $T_{a_1\cdots a_N}$ can be reduced into a network comprised by $N_{\rm T}$ number of tensors $t_{b_1\cdots b_n}$ with less rank, namely, $n<N$, and for simplicity, we require each index $b_j$ takes the same $q$ number of values.

Note that $|\Psi\rangle$ can also be written into the Euclidean path integral form as~\cite{Casini:2009sr}
\be\label{pathintT}\Psi[\phi(x)]=\mathcal{N}^{-\frac{1}{2}}\int\prod_{0<\tau<\infty}D\phi(\tau,x)\delta
\left(\phi(0,x)-\phi(x)\right)e^{-I_{\rm E}[\phi]},
\ee
where $\phi(x)$ are fields, $I_{\rm E}[\phi]$ is the Euclidean action of the manybody system and $\mathcal{N}^{\frac{1}{2}}$ is the normalization factor, respectively.

On the other hand, the wave function $\Psi_{\rm G}[h_{IJ},\varphi]$ of a spacetime can also be expressed as the Euclidean path integral~\cite{Hartle:1983ai}
\be\label{pathintGr}\Psi_{\rm G}[h_{IJ},\varphi]=\int_C D[g]D[\varphi]e^{-I_{\rm E}[g,\varphi]},
\ee
where $C$ indicates a class of spacetimes with compact boundary such that the reduced metric $h_{IJ}$ and matter fields $\varphi$ satisfy the given boundary conditions.

Recall that the fundamental equation of the AdS/CFT correspondence is the equivalence between the partition function (generating functional) of the bulk gravity and that of the boundary CFT~\cite{Maldacena:1997re,Gubser:1998bc,Witten:1998qj}
\be\label{adscft}Z_{{\rm AdS}}=Z_{\rm CFT}.
\ee
For general spacetime backgrounds, the holographic principle indicates that the partition function of the bulk theory should equal to that of the boundary theory. Furthermore, the wave functionals $\Psi[\phi(x)]$ and $\Psi_{\rm G}[h_{IJ},\varphi]$ contain all of the informations of their corresponding systems respectively and they will reduce to the associated partition functions when the initial states are chosen as the $\delta$ function source. Therefore, if a quantum manybody system is holographically dual to a gravitational theory living in higher dimensional spacetime, we conjecture that the relation
\be\label{holowave}\Psi[\phi(x)]=\Psi_{\rm G}[h_{IJ},\varphi]
\ee
is held. Eq.(\ref{holowave}) can be viewed as a generalization of eq.(\ref{adscft}), and it is a bridge to connect the dynamics of the two sides.

\section{Transformation of d.o.f. from the tensor network to the metric}
If a tensor network can describe gravity, a crucial question to ask is how are the d.o.f. of the former mapping to those of the latter, similar to the relation of field/operator duality in the gauge/gravity duality. Interestingly, we found that the entanglement entropy obtained from the tensor network plays a vital important role. To see this, note that the open indices of a tensor network correspond to the physical d.o.f. of the quantum manybody system~\cite{Orus:2013kga}, for an arbitrary given region (without open indices inside it) in the tensor network with volume $V$ (denoted as region A) and boundary $\partial V$, the number of external indices (legs) $m$ of the region are proportional to its boundary area $\Sigma_A$, when each leg has $q$ number of excitations, there are $q^m$ microstates on $\Sigma_A$, and hence the associated entropy is $S_A=k_B\ln q^m=m k_B\ln q\propto m\propto \Sigma_A$, which just describes the entanglement entropy between regions $V$ and $\bar V$ (denoted as region B) and obeys the area law.

Moreover, assume that the tensor network forms a $d-1$-dimensional flat space (note that this space is not necessarily be the real space), the entanglement renormalization method together with the holographic entanglement entropy proposal suggest that $S_A$ is a quarter of the area $A$ of a bulk co-dimensional-2 minimal surface $\lambda_A$ with boundary $\partial V$
\be\label{Sa}S_A=\frac{A}{4 l_p^{d-1}},
\ee
where $A=\int \sqrt{\tilde{\lambda}}d^{d-1}\xi$, with $\tilde{\lambda}_{ij}$ and $\xi$ the reduced metric and coordinate on $\lambda_A$, and $l_p$ is the fundamental length scale of the $d+1$-dimensional spacetime. We regard this geometry as the one emerged from the tensor network. Nevertheless, for dimensional analysis, it is expected that $l_p^{d-1}$ be the gravitational coupling constant $G_{d+1}$. On the other hand, $S_A$ is calculated from the von Neumann entropy
\be\label{von Neumann}S_A=-{\rm tr}\rho_A\ln\rho_A,
\ee
where $\rho_A$ is the reduced density matrix of the subsystem A, which can be expressed as
\be\label{rhoa} \rho_A &=&\int D\phi_B\Psi^*[\phi_B\oplus\phi_A]\Psi[\phi_B\oplus\phi'_A]\nno\\
&=&\frac{1}{\mathcal{N}}\int_{\phi(\vec{x},0^-)=\phi'_A}^{\phi(\vec{x},0^+)=\phi_A}D\phi e^{-I_{\rm E}[\phi]},
\ee
in which $\phi_B$ are fields belonging to the region B and ${\rm tr}\rho_A=1$.

An important hint from eq.(\ref{Sa}) is that it governs the transformation of the d.o.f. from the tensor network to the reduced geometry on $\lambda_A$, namely, $\tilde{\lambda}_{ij}$. In addition, $\tilde{\lambda}_{ij}$ is obtained from the reduced geometry $\tilde{h}_{IJ}$ on a time slice $\Sigma_t$ via $\tilde{\lambda}_{ij}=\frac{\partial y^I}{\partial \xi^i}\frac{\partial y^J}{\partial \xi^j}\tilde{h}_{IJ}$ (for static minimal surface), where $y^I$ is coordinate on $\Sigma_t$. Consequently, the d.o.f. of the tensor network are transformed to those of reduced geometry $\Sigma_t$, namely, $\tilde{h}_{IJ}$, which indicates the similar relation with eq.(\ref{holowave})
\be\label{holowave2}\Psi[\phi(x)]=\Psi[\phi[\tilde{h}_{IJ}]],
\ee
since the wave functional $\Psi[\phi(x)]$ is a scalar function, eq.(\ref{holowave2}) is just a change of variables in the wave functional. Clearly, if $\tilde{h}_{IJ}$ can represent the real spacetime geometry, eq.(\ref{holowave2}) and eq.(\ref{holowave}) are the same.

To see more clearly how the d.o.f. of the tensor network and the emerged geometry are connected with each other, let us consider variation on both sides. From eqs.(\ref{Sa})(\ref{von Neumann}), we have
\be\label{variationS}\delta S_A=-{\rm tr}\left(\delta\rho_A\ln\rho_A \right)=-\frac{1}{8G_{d+1}}\int \sqrt{\tilde{\lambda}}d^{d-1}\xi\tilde{\lambda}_{ij}\delta\tilde{\lambda}^{ij}.
\ee
When the variation is caused by a single operator perturbation, namely,
\be\label{perturbation}I^{(0)}_{\rm E}\rightarrow I_{\rm E}=I^{(0)}_{\rm E}+g_s\int d^dx \mathcal{O}(x),
\ee
where $g_s$ is the coupling constant, the density matrix changes as
\be\label{deltarho}\delta\rho_A &=&\rho_A-\rho_A^{(0)}\nno\\
&=&\frac{1}{\mathcal{N}}\int_{\phi(\vec{x},0^-)=\phi'_A}^{\phi(\vec{x},0^+)=\phi_A}D\phi e^{-I^{(0)}_{\rm E}-g_s\int d^dx \mathcal{O}(x)}-\frac{1}{\mathcal{N}^{(0)}}\int_{\phi(\vec{x},0^-)=\phi'_A}^{\phi(\vec{x},0^+)=\phi_A}D\phi e^{-I^{(0)}_{\rm E}},
\ee
where
\be \mathcal{N}&=&\int D\chi\int_{\phi(\vec{x},0^-)=\phi(\vec{x},0^+)=\chi}D\phi e^{-I^{(0)}_{\rm E}-g_s\int d^dx \mathcal{O}(x)}\nno\\
&=& \mathcal{N}^{(0)}\left(1-g_s\int d^dx\langle\mathcal{O}(x)\rangle +\frac{g_s^2}{2}\int d^dx\int d^dx'\langle\mathcal{O}(x)\mathcal{O}(x')\rangle +...\right),
\ee
then at the first order perturbation,
\be\label{deltarho1} \delta\rho_A=g_s\rho_A^{(0)}\left(\int d^dx\langle\mathcal{O}(x)\rangle-\int d^dx\mathcal{O}(x)\right).
\ee
Substituting eq.(\ref{deltarho1}) into eq.(\ref{variationS}), one then obtain the explicit relationship between $\mathcal{O}(x)$ and $\delta\tilde{\lambda}^{ij}$.

\section{Emergence of the gravitational dynamics}
In order that the emergent reduced metric $\tilde{h}_{IJ}$ can describe the real spacetime geometry, it needs to satisfy the gravitational dynamical equation, namely, Einstein's equation or its equivalent form. Considering a real $d+1$-dimensional stationary spacetime with
\be ds^2=-N^2 dt^2+h_{IJ}(N^I dt+dx^I)(N^J dt+dx^J),\ee
the Hamiltonian formalism gives the Hamiltonian and the momentum constraints, and in the canonical quantization, they become the Wheeler-DeWitt equation and the quantum momentum constraint equation that the wave function of the spacetime to satisfy~\cite{DeWitt:1967yk}
\be\label{WD}\left\{-G_{IJLK} \frac{\delta^2}{\delta h_{IJ}\delta h_{KL}}-\sqrt{-h} \left(^{(d)}R -2\Lambda\right)\right\}\Psi[h_{IJ}]&=&0,\\
\left\{\frac{\delta}{\delta h_{IJ}}\Psi[h_{IJ}]\right\}_{|I} &=&0,
\ee
where $G_{IJKL}=h^{-1/2}\left(\frac 1 2 \left(h_{IK}h_{JL}+h_{IL}h_{JK}\right)-\frac{1}{d-1}h_{IJ}h_{KL}\right)$ is the supermetric and $^{(d)}R$ is $d$-dimensional Ricci curvature constructed from the reduced metric $h_{IJ}$, and here we only consider the bulk to be the vacuum, namely, without the matter fields.

Furthermore, the ground state wave function $\Psi[\varphi(x)]$ described by a tensor network satisfies the Schr\"{o}dinger equation $\mathcal{\hat{H}}\Psi[\varphi(x)]=0$. Therefore, from eqs.(\ref{holowave2}) and (\ref{holowave}), if a tensor network can describe a real spacetime, its associated Schr\"{o}dinger equation should be able to rewritten in the same form as the Wheeler-DeWitt equation, i.e.
\be\label{holoWD}\mathcal{\hat{H}}\Psi[\varphi(x)]=\left\{-\tilde{G}_{IJKL} \frac{\delta^2}{\delta \tilde{h}_{IJ}\delta \tilde{h}_{KL}}-\sqrt{-\tilde{h}} \left(^{(d)}\tilde{R} -2\tilde{\Lambda}\right)\right\}\Psi[\varphi[\tilde{h}_{IJ}]]&=&0,\\
\left\{\frac{\delta}{\delta \tilde{h}_{IJ}}\Psi[\varphi[\tilde{h}_{IJ}]]\right\}_{|I} &=&0,
\ee
which means that $\tilde{h}_{IJ}$ can describe a real spacetime metric $h_{IJ}$, where $\mathcal{\hat{H}}$ is the Hamiltonian density of quantum manybody system.

\section{Projecting the Wheeler-DeWitt equation on the AdS boundary}
In the original holographic approach, the CFT or QFT is located at the asymptotical spatial boundary. Therefore, the Hamiltonian $\mathcal{\hat{H}}$ of the tensor network should be expressed as an operator in terms of variables on the spatial boundary. This can be done by projecting the Wheeler-DeWitt equation (\ref{WD}) on the AdS boundary. Let's considering the $d+1$-dimensional static AdS spacetime with metric
\be\label{staticads} ds^2&=&=g_{AB}dX^A dX^B=-N^2(r)dt^2+h_{IJ}(r)dy^I dy^J\nno\\
&=&-N^2(r)dt^2+h_{rr}(r)dr^2+h(r)dx_i^2,
\ee
where the asymptotical spatial boundary $\Sigma_r$ (with induced metric $\gamma_{\alpha\beta}$ and coordinate $x^\alpha$) is located at $r\rightarrow \infty$. Besides, denoting the spatial boundary of the time slice $\Sigma_t$ to be $\mathcal{B}_t$ (with induced metric $\sigma_{ab}$ and coordinate $\theta^a$), which is the region of taking $\Sigma_t$ to $r\rightarrow\infty$. The Ricci tensor $^{(d)}R_{IJ}=-(d-1)h_{IJ}/L^2$, which gives $^{(d)}R-2\Lambda=0$ for static AdS spacetime case, where $L$ is the curvature radius of the AdS spacetime. Then the Wheeler-DeWitt equation (\ref{WD}) becomes
\be\label{WDstatic}\left\{-G_{IJLK} \frac{\delta^2}{\delta h_{IJ}\delta h_{KL}}\right\}\Psi[h_{IJ}]&=&0.
\ee
Furthermore, the displacement on $\Sigma_r$ is
\be dX^A=\frac{\partial X^A}{\partial t}dt+\frac{\partial X^A}{\partial \theta^a}d\theta^a=Nn^A dt+e^A_a d\theta^a,
\ee
where $n^A$ is the timelike unit normal vector of $\Sigma_t$, then the reduced line-element on $\Sigma_r$ is
\be ds^2&=&g_{AB}(Nn^A dt+e^A_a d\theta^a)(Nn^B dt+e^B_b d\theta^b)\nno\\
&=&-N^2 dt^2+g_{AB}e^A_a e^B_b d\theta^a d\theta^b\nno\\
&=&-N^2 dt^2+\sigma_{ab}d\theta^a d\theta^b\nno\\
&\equiv&\gamma_{\alpha\beta}dx^\alpha dx^\beta,
\ee
which gives $\sqrt{-\gamma}=N\sqrt{\sigma}$. In addition, $h_{IJ}$ and $\sigma_{ab}$ are related by $\sigma_{ab}=h_{IJ}\frac{\partial y^I}{\partial\theta^a}\frac{\partial y^J}{\partial\theta^b}\equiv h_{IJ}e^I_a e^J_b$, then $\frac{\delta}{\delta h_{IJ}}=e^I_a e^J_b \frac{\delta}{\delta\sigma_{ab}}$. Therefore, the Wheeler-DeWitt equation (\ref{WD}) can be rewritten as
\be\label{WDads}&&\left\{-G_{IJLK} \frac{\delta^2}{\delta h_{IJ}\delta h_{KL}}\right\}\Psi[h_{IJ}]\nno\\
&&=\frac{1}{\sqrt{h}}\left(\frac 1 2 \left(h_{IK}h_{JL}+h_{IL}h_{JK}\right)-\frac{1}{d-1}h_{IJ}h_{KL}\right)e^I_a e^J_b e^K_c e^L_d\frac{\delta^2}{\delta\sigma_{ab}\delta\sigma_{cd}}\Psi[\sigma_{ab}]\nno\\
&&=\frac{1}{\sqrt{h}}\left(\frac 1 2 \left(\sigma_{ac}\sigma_{bd}+\sigma_{ad}\sigma_{bc}\right)-\frac{1}{d-1}\sigma_{ab}\sigma_{cd}\right)
\frac{\delta^2}{\delta\sigma_{ab}\delta\sigma_{cd}}\Psi[\sigma_{ab}]\nno\\
&&=0.
\ee
While at fixed time, the reduced metric $\gamma_{\alpha\beta}$ on the AdS boundary reduces to $\sigma_{ab}$, namely, $\gamma_{ab}=\sigma_{ab}=h_{IJ}e^I_a e^J_b$. Consequently, eq.(\ref{WDads}) is equivalent to
\be\label{WDadsbdy}
\frac{1}{\sqrt{h}}\left(\frac 1 2 \left(\gamma_{ac}\gamma_{bd}+\gamma_{ad}\gamma_{bc}\right)-\frac{1}{d-1}\gamma_{ab}\gamma_{cd}\right)
\frac{\delta^2}{\delta\gamma_{ab}\delta\gamma_{cd}}\Psi[\gamma_{ab}]=0,
\ee
which is the Wheeler-DeWitt equation on the AdS boundary.

According to our proposal, in order that a tensor network can generate a real AdS spacetime, its associated Schr\"{o}dinger equation should be expressed as
\be\label{holoWDbdy}\mathcal{\hat{H}}\Psi[\varphi(x)]=\frac{1}{\sqrt{\tilde{h}}}\left(\frac 1 2 \left(\tilde{\gamma}_{ac}\tilde{\gamma}_{bd}+\tilde{\gamma}_{ad}\tilde{\gamma}_{bc}\right)-
\frac{1}{d-1}\tilde{\gamma}_{ab}\tilde{\gamma}_{cd}\right)
\frac{\delta^2}{\delta\tilde{\gamma}_{ab}\delta\tilde{\gamma}_{cd}}\Psi[\tilde{\gamma}_{ab}]=0.
\ee
Clearly, not every tensor network can satisfy eq.(\ref{holoWDbdy}) and it is nontrivial to find out concrete examples of quantum manybody systems (concrete $\mathcal{\hat{H}}$) which can describe gravity. Besides, it is known that not all of the wave functions of CFTs are dual to classical gravity geometry. A fundamental question is what kinds of conditions a wave function must satisfy in order to have a classical gravity duality. We propose that the answers lie in the Wheeler-DeWitt equation.



\section{Conclusions and Discussions}\label{conclusion}
The gauge/gravity dualities have provided us very powerful tools to study the CFT or QFT from their dual gravitational theories in the bulk. However, the study in the inverse direction, i.e. from boundary to the bulk, is not so clear and straightforward. The essential questions are how to construct the bulk geometry, and especially, how to construct or generate the bulk gravitational dynamics from the information of the boundary QFT. In this paper, we studied the bulk reconstruction of the AdS spacetime from tensor networks on the boundary and proposed a novel approach to generate the bulk gravitational dynamics by combining the ideas of the holographic entanglement entropy and the canonical quantization of quantum gravity and argue that there is a connection between the boundary Schr\"{o}dinger equation and the Wheeler-DeWitt equation in the bulk gravity side. Our approach deepens the understanding of the gauge/gravity duality and makes its formalism more complete. Beside, the approach also support the emergent picture of gravity. There remains a lot of works to do, such as finding (or constructing) appropriate tensor networks models to generate the desired gravitational backgrounds and extending our method to stationary spacetime cases.

\section*{Acknowledgement}
We would like to thank Ling-Yan Hung and Rong-xin Miao for collaboration at the initial stage of this work, and thank Yi Ling for useful discussions. JRS was supported by the National Natural Science Foundation of China under Grant No.~11675272, YS was supported by by China Postdoctoral Science Foundation (No. 2019M653137).




\end{document}